\documentclass[%
reprint,
superscriptaddress,
showpacs,%
nofootinbib,
nobibnotes,
amsmath,amssymb,
aps,
pre,
longbibliography,
floatfix,
]{revtex4-1}

\usepackage[utf8]{inputenc}
\usepackage[T1]{fontenc}
\usepackage[english]{babel}
\usepackage{amsfonts}
\usepackage{amsmath}
\usepackage{graphicx}
\usepackage{color}
\usepackage{hyperref}

\hypersetup{colorlinks = true, allcolors = blue}

\newcommand{\widfig}[2]{\includegraphics[width=#1\columnwidth]{#2}}
\newcommand{\fulfig}[1]{\includegraphics[width=\textwidth]{#1}}

\newcommand{\mat}[1]{\mbox{\boldmath{$\mathrm{#1}$}}}
\newcommand{\sspace}[1]{\mathbb{#1}}
\newcommand{\spanspc}{\mathop{\mathrm{span}}}

\newcommand{\identmat}{\mat I}
\newcommand{\diagmat}{\mathop{\mathrm{diag}}}
\newcommand{\transp}{\mathrm{T}}
\newcommand{\maxval}{\mathop{\mathrm{max}}}
\newcommand{\minval}{\mathop{\mathrm{min}}}

\newcommand{\phas}{P\lowercase{h}}
\newcommand{\fsyn}{FS}
\newcommand{\phc}{\text{p}}
\newcommand{\fsc}{\text{f}}
\newcommand{\spr}{\text{s}}

\newcommand{\clutim}{\mathcal{T}}
\newcommand{\clvpn}{p}

\begin{document}

\title{Predictable nonwandering localization of covariant Lyapunov
  vectors and cluster synchronization in scale-free networks of
  chaotic maps}

\author{Pavel V. Kuptsov}\email[Corresponding author. Electronic
address:]{p.kuptsov@rambler.ru}%
\affiliation{Institute of electronics and mechanical engineering, Yuri
  Gagarin State Technical University of Saratov, Politekhnicheskaya
  77, Saratov 410054, Russia}%

\author{Anna V. Kuptsova}%
\affiliation{Institute of electronics and mechanical engineering, Yuri
  Gagarin State Technical University of Saratov, Politekhnicheskaya
  77, Saratov 410054, Russia}%




\date{\today}

\begin{abstract}
  Covariant Lyapunov vectors for scale-free networks of H\'{e}non maps
  are highly localized. We revealed two mechanisms of the localization
  related to full and phase cluster synchronization of network
  nodes. In both cases the localization nodes remain unaltered in
  course of the dynamics, i.e., the localization is
  nonwandering. Moreover this is predictable: the localization nodes
  are found to have specific dynamical and topological properties and
  they can be found without computing of the covariant vectors. This
  is an example of explicit relations between the system topology, its
  phase space dynamics, and the associated tangent space dynamics of
  covariant Lyapunov vectors.
\end{abstract}

\maketitle

\section{Introduction} 

Localization properties of Lyapunov vectors in spatio-temporal chaotic
systems attract a permanent interest since the early works till the
present
days~\cite{Kaneko1986436,GiaPol91,FalMarVul91,Morris2012}. Recently it
has been renewed due to the discovery of algorithms for covariant
Lyapunov vectors (CLVs)~\cite{CLVGin,CLVWS}. The evolution of these
vectors is governed by linear equations under chaotic forcing, so that
their localization can be treated as a sort of Anderson
localization~\cite{GiaPol91}. The localization sites indicate unstable
areas of a system, that, in particular, is important for atmosphere
dynamics prediction~\cite{BuizPal95}. For homogeneous systems the
localization sites of the covariant vectors wander irregularly so that
their dynamics can be described by stochastic equation of
Kardar-Parisi-Zhang~\cite{PikKur94,*PikPol98,CLVPazo}. In contrast,
the localization positions in inhomogeneous systems are pinned at
certain fixed positions~\cite{SzendLopRodr08}.

In this paper we analyze properties of CLVs for scale-free networks of
chaotic maps. We show that due to the presence of cluster
synchronization the CLVs are localized. The first mechanism of the
localization is related to the full synchronization clusters, and
second one appears due to the existing of large phase synchronized
clusters. Both of the localizations are nonwandering, i.e., nonzero
sites of the vectors remain unchanged in course of the
dynamics. Moreover these nodes have specific topological and dynamical
properties so that they can be identified without computing the CLVs.
This is an example of explicit relations between the system topology,
its phase space dynamics, and the associated tangent space dynamics of
CLVs.

The paper is organized as follows. In Sec.~\ref{sec:model} we
introduce the considered network and discuss its
dynamics. Section~\ref{sec:tanspc} describes the structure of the
tangent space of the network. The mechanism of CLVs localization on
clusters of full synchronization is described in Sec.~\ref{sec:nwlfs},
and in Sec.~\ref{sec:nwlph} we discuss the localization related to
phase clusters. Finally, Sec.~\ref{sec:summ} summarizes the paper
results.

\section{\label{sec:model}Model system and 
cluster synchronization}

\subsection{Dynamical network equations and network structure}

We consider a network of H\'{e}non maps build
as a generalization of the H\'{e}non chain from Ref.~\cite{PolTor92a}:
\begin{gather}
  \begin{aligned}\label{eq:netw_henon}
    x_n(t+1)&=\alpha-[x_n(t)+\epsilon h_n(t)]^2+y_n(t),\\
    y_n(t+1)&=\beta x_n(t),
  \end{aligned}\\
  \label{eq:gn}
  h_n(t)=\sum_{j=1}^N \frac{a_{nj}}{k_n}x_j(t)-x_n(t),\;
  k_n=\sum_{j=1}^Na_{jn},
\end{gather}
where $N$ is the number of network nodes, $t=0,1,2\ldots$ is discrete
time, $a_{nj}\in \{0,1\}$, $a_{nn}=0$ are the elements of the $N\times
N$ adjacency matrix $\mat A$, and $k_n$ is degree of the $n$th node,
i.e., the number of its connections. $\alpha=1.4$ and $\beta=0.3$ are
the parameters, controlling local dynamics, and $\epsilon\in [0,1]$ is
the coupling strength. The system is time-reversible: $x_n(t)=y_n(t+1)/\beta$,
$y_n(t)=-\alpha+[y_n(t+1)+\epsilon h'_n(t+1)]^2/\beta^2+x_n(t+1)$,
where $h'_n(t)=\sum_{j=1}^N \frac{a_{nj}}{k_n}y_j(t)-y_n(t)$.

We consider random networks with scale-free structure generated via a
stochastic process described in Ref.~\cite{ScaleFreeNetw}. The process
starts from two linked nodes. At each iteration we add one node to the
network and one link connecting it with one of the existing nodes. The
node to connect is chosen at random with probability that is
proportional to its connectivity degree $k_n$, i.e, via so called
preferential attachment mechanism. After $N-1$ steps we obtain a
network with $N$ nodes and $N-1$ connections. The node degree
distribution for such networks has a power law shape $P(k)\sim
k^{-3}$.  An example of the network is shown in Fig.~\ref{fig:netw1}
(this figure is discussed in detail below).

By construction, the networks under consideration do not have
loops. It means that starting from any node one can not return to it
without moving back. The networks always have a lot of star-like
structures when one hub node is connected with many subordinate ones,
like, for example, node 10 in Fig.~\ref{fig:netw1}. Moreover these
structures can form a hierarchy, see the hub node 11 that is
subordinate with respect to node 10. The structure of considered
networks is essentially inhomogeneous. Usually a few nodes are
connected with very many others, and many nodes have only one
link. All of these properties are found to result in a very long
transient time required for the network to arrive at stationary
regime. This will be discussed in Sec.~\ref{sec:conv}.

\subsection{The largest Lyapunov exponent}

\begin{figure}
  \begin{center}
    \widfig{1}{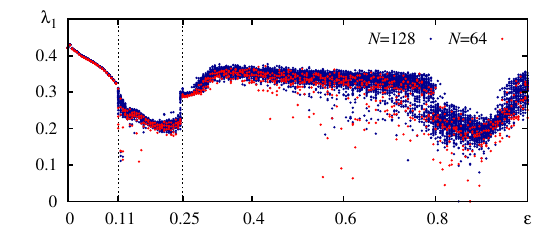}
  \end{center}
  \caption{\label{fig:lyeps}(color online) The first Lyapunov exponent
    vs. $\epsilon$ for $N=128$ and $64$. Each point is computed
    independently with a new matrix $\mat A$ and initial
    conditions. At $N=128$ and $64$ there are 50 and 5 points,
    respectively, for each $\epsilon$. Lines at $\epsilon=0.11$ and
    $0.25$ delimit the area of interest.}
\end{figure}

The dynamics of the network~\eqref{eq:netw_henon} is, in general,
chaotic. To characterize it we compute Lyapunov exponents using the
standard algorithm suggested in Refs.~\cite{Benettin,Shimada79} (see
also Ref.~\cite{CLV2012} for a review).

Figure~\ref{fig:lyeps} shows the largest Lyapunov exponent $\lambda_1$
at different coupling strengths. At $\epsilon<0.11$ the exponent
unambiguously depends on $\epsilon$ regardless of the network matrix
$\mat A$, initial conditions, and the network size. This occurs
because the nodes interacts weakly with each others, so that the
detailed network structure is not very important. The nodes within
this area do not demonstrate any concerted oscillations. The area
$0.11<\epsilon<0.25$ clearly differs from all others. The dependence
$\lambda_1(\epsilon)$ is ambiguous here: every new combination of the
network matrix $\mat A$ and initial conditions are characterized with
their own $\lambda_1$. Another feature of this area is lower values of
$\lambda_1$ with respect to the surrounding areas. This is due to the
cluster synchronization emerging here, see the discussion below in
Sec.~\ref{sec:clusters}. The dependence $\lambda_1(\epsilon)$ remain
ambiguous at $\epsilon>0.25$, though the exponents becomes higher. At
$\epsilon>0.8$ the exponents again becomes lower so that this area is
similar to the marked area $0.11<\epsilon<0.25$.

In what follows we shall restrict ourselves with the area
$0.11<\epsilon<0.25$.

\subsection{\label{sec:clusters}Full and phase cluster synchronization}

Though the synchronization of the whole network is not observed, the
nodes can form clusters of synchronized oscillations. Both full and
phase synchronization is possible. The former stands for the
equivalence of variables at the synchronized nodes, and the latter
implies the coincidence of positions of minima and maxima of
synchronized time series. The fully synchronized nodes will be
referred to as \fsyn-clusters, and phase synchronized nodes will be
called \phas-clusters.

The phase cluster synchronization of networks nodes is studied in
Ref.~\cite{JalanAmritkar2003,*JalanAmritkar2005}. According to the
approach suggested there, one can detect the \phas-clusters computing
phase distances. Given a starting time $t_0$ and a time interval
$\clutim$, count at $t_0\leq t<t_0+\clutim$ the numbers $\nu_m$ and
$\nu_n$ of local minima of $x_m(t)$ and $x_n(t)$, respectively, and
also find the number $\nu_{mn}$ of simultaneous minima of $x_m$ and
$x_n$. Then the phase distance is computed as
\begin{equation}
  \label{eq:phase_dist}
  d_{mn}=1-\nu_{mn}/\maxval(\nu_m,\nu_n).
\end{equation}
When it vanishes all the minima of $x_m$ and $x_n$ occur
simultaneously and this is the case of phase synchronization of $m$th
and $n$th nodes over the time interval $\clutim$. To identify the
\phas-clusters one can build an auxiliary graph whose $n$th and $m$th
nodes are linked if $d_{mn}=0$ and find the clusters as connected
components of this graph.

Nonzero $d_{mn}$ is a fraction of time when the nodes $m$ and $n$ are
not synchronized. Thus the minimum of $d_{mn}$ over $n$, i.e.,
\begin{equation}
  \label{eq:phase_desyn}
  \tilde d_m=\minval\{d_{mn}|n=1\ldots N\},
\end{equation}
can be treated as degree of the desynchronization of the $m$th node
with the rest of the network.

The \fsyn-clusters can be identified using the matrix of mean
absolute differences between dynamical variables over the computation
interval $\clutim$:
\begin{equation} 
  \label{eq:ident_dist}
  q_{mn}=\sum_{t=0}^{\clutim-1} |x_m(t_0+t)-x_n(t_0+t)|/\clutim
\end{equation}
The \fsyn-clusters correspond to connected components of an auxiliary
graph whose $m$th and $n$th nodes are connected when $q_{mn}=0$. In
actual numerical simulations we considered two nodes as synchronized
if $q_{mn}<10\epsilon_m$, where $\epsilon_m\approx 10^{-16}$ is the
machine epsilon for double precision variables that was employed.

The length of the interval $\clutim$ for which the cluster detection
is performed can influence the resulting picture. As we discuss in
this section below and in Sec.~\ref{sec:conv}, there exist so called
floating nodes that intermittently can either belong to one of the
\phas-clusters or oscillate separately. With a large $\clutim$ we
consider clusters including only permanent nodes, while performing a
serial cluster detections with a small $\clutim$ we can take into
account fluctuations arising due to the floating nodes.

\begin{figure}
  \begin{center}
    \widfig{1}{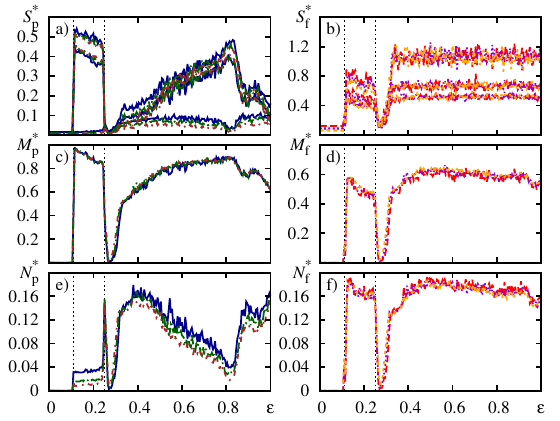}
  \end{center}
  \caption{\label{fig:clustering}(color online) (a,b) Rescaled sizes
    of three largest \phas- and \fsyn-clusters, see
    Eq.~\eqref{eq:scale_s}. (c,d) Rescaled numbers of nodes attached
    to all \phas- and \fsyn-clusters, see
    Eq.~\eqref{eq:scale_m}. (e,f) Rescaled numbers of \phas- and
    \fsyn-clusters, see Eq.~\eqref{eq:scale_n}. All values are
    averaged over 25 computations with different matrices $\mat A$ and
    initial conditions at each $\epsilon$. $\clutim=10000$. Different
    curves in each panel correspond to $N=62$, $128$, and
    $256$. Vertical doted lines are plotted at $\epsilon=0.11$ and
    $0.25$ to delimit the area of interest.}
\end{figure}

Figure~\ref{fig:clustering} illustrates the cluster synchronization of
networks with $N=64$, $128$, and $256$ nodes that is observed at
different $\epsilon$. Panels (a) and (b) show rescaled sizes
\begin{equation}
  \label{eq:scale_s}
  S^*_\phc=S_\phc/N, \; S^*_\fsc=S_\fsc/\sqrt{N}
\end{equation}
of three largest \phas- and \fsyn-clusters, respectively. Panels (c)
and (d) represents rescaled numbers
\begin{equation}
  \label{eq:scale_m}
  M^*_\phc=M_\phc/N, \; M^*_\fsc=M_\fsc/N
\end{equation}
of nodes attached to all \phas- and \fsyn-clusters,
respectively. Panels (e) and (f) show rescaled numbers
\begin{equation}
  \label{eq:scale_n}
  N^*_\phc=N_\phc/N, \; N^*_\fsc=N_\fsc/N
\end{equation}
of \phas- and \fsyn-clusters, respectively. The clusters appears at
$\epsilon=0.11$. As one can see in panel (a) in the area
$0.11<\epsilon<0.25$ there are two large \phas-clusters whose relative
sizes are $S^*_\phc\approx 0.4\div 0.5$. The curves in panel (e)
plotted for different $N$ do not coincide, but not rescaled curves
$N_\phc$ do so (not shown), i.e., the number of \phas-clusters does
not depend on $N$. Since $S_\phc\sim N$, see Eq.~\eqref{eq:scale_s},
regardless of $N$ these clusters includes the bulk of nodes. However,
as follows from panel (c) and Eq.~\eqref{eq:scale_m}, the
\phas-clusters includes at any $N$ approximately 85\% of nodes, so
that always there are nodes not synchronized with \phas-clusters.

Despite the \phas-clusters, the number of \fsyn-clusters scales as
$N_\fsc\sim N$ and also the total number of nodes attached to all
\fsyn-cluster grows as $M_\fsc\sim N$. It presumes that the mean size
of \fsyn-nodes is constant. However the size of the largest cluster
grows: at $N=64$, $128$, and $256$ the sizes are $S_\fsc\approx 6$,
$9$, and $13$, respectively. According to Eq.~\eqref{eq:scale_s}, the
sizes of the largest \fsyn-clusters scales with $N$ as $S_\fsc\sim
\sqrt N$.

At the right boundary of the discussed area at $\epsilon=0.25$ the
large \phas-clusters desintegrate into many small ones, see the spike
of $N^*_\phc$ in the panel (e). Moreover, in this area $N_\phc$ starts
to scale as $N_\phc\sim N$. As $\epsilon$ further grows all clusters
disappears but then their number again increase. Notice the identical
behaviour of curves in panels (c,e) and (d,f), respectively, around
$\epsilon\approx 0.3$. It indicates the presence here of
\fsyn-clusters only. Subsequent growth of $\epsilon$ results in
reappearing of the \phas-clusters, but their number is still high. At
$\epsilon\approx 0.4$ the number of \phas-clusters starts to decay,
panel (e), and the number of the attached nodes increases, panel
(c). Also observe the growth of the first two largest clusters, panel
(a). As for the \fsyn-clusters, their sizes, panel (b), the number of
attached nodes, panel (d), and their total number, panel (f), remains
approximately unchanged. At $\epsilon\approx 0.8$ one again observes
the situation when there are two large \phas-clusters and many small
\fsyn-clusters. But contrary to the area $0.11<\epsilon<0.25$, this
area is much narrower and when $\epsilon$ gets larger the
desintegration of \phas-clusters occurs within the wider range of
$\epsilon$.

As already mentioned above, we shall consider the dynamics of the
network within the area at $0.11<\epsilon<0.25$.

\begin{figure}
  \begin{center}
    \widfig{1}{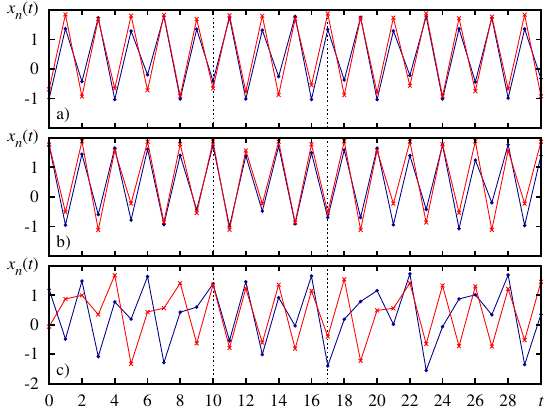}
  \end{center}
  \caption{\label{fig:sptm}(a,b) Oscillations at nodes belonging to
    two large \phas-clusters. (c) Separated nodes not synchronized
    with others. Vertical dotted lines delimit the interval when both
    separated nodes in panel (c) are attached to the cluster
    represented in panel (b). $N=128$, $\epsilon=0.17$.}
\end{figure}

Figure~\ref{fig:sptm} illustrates behaviour of synchronized and
separated nodes, panels (a,b) and (c), respectively, within the area
of interest, when almost all nodes belong to two large
\phas-clusters. Observe in panels (a) and (b) strict alternations of
maxima and minima of variables attached to \phas-clusters and
irregular variations of their amplitudes. Also compare the panels (a)
and (b): the oscillations of \phas-clusters have opposite phases. The
separated nodes, panel (c), oscillate irregularly, however for some
time they can be attached to one of the clusters, see area $25<t<30$
in panel (c).

If a node spends an essential part of time being synchronized with
others however can lose intermittently the synchronization, it will be
called a floating, according to the notation suggested in
Ref.~\cite{JalanAmritkar2003,*JalanAmritkar2005}.

\subsection{\label{sec:conv}Convergence of the cluster structure}

\begin{figure}
  \begin{center}
    \widfig{1}{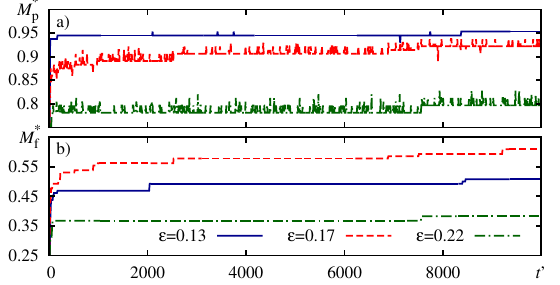}
  \end{center}
  \caption{\label{fig:conv}Convergence of the (a) \phas- and (b)
    \fsyn-clusters. $t'$ enumerates the cluster detection procedures
    performed over the intervals $\clutim=100$ in course of the
    evolution of the system. $N=128$, $\epsilon=0.13$, $0.17$, and
    $0.22$.}
\end{figure}

The network~\eqref{eq:netw_henon} converges very slowly to its
stationary regime. As one can see in Fig.~\ref{fig:conv}, the relative
numbers $M^*_\phc$ and $M^*_\fsc$ of nodes attached to \phas- and
\fsyn-clusters, respectively, can change even after a very long
evolution time. Since in this figure the clusters are identified over
the intervals $\clutim=100$, the total evolution time of the system is
$t=10^6$. The represented examples are not very typical in a sense
that we tried approximately ten different matrices $\mat A$ and
initial conditions for each $\epsilon$ to show the cases with the
worst convergence. However, the convergence in other cases is not much
faster. Nevertheless, both in Fig.~\ref{fig:conv} and in all other
cases we tried the curves always behaved as if they approached to
limiting values. Thus we can conjecture that the stationary regime
exists and take a long transient time to approach it,
$t_{\text{trans}}=5\times 10^5N/64$.

Observe frequent peaks and dips on the curves for $M^*_\phc$, see
Fig.~\ref{fig:conv}(a). They appear due to the floating nodes that
intermittently attach and detach the \phas-clusters. The floating
nodes exist only with respect to \phas-clusters; if a node gets
attached to a \fsyn-cluster it stays synchronized permanently, see
Fig.~\ref{fig:conv}(b). 

Curves in Fig.~\ref{fig:conv}(a) can be treated as a highly
fluctuating signal. However, the observed fluctuations appear due to
the serial cluster detection with sufficiently short $\clutim$. One
can change the definition of observable variables and perform the
clusters detected just once over the whole computation time. The
clusters defined in this way are stationary, but also there are non
cluster nodes oscillating chaotically. Below we shall employ both
approaches.

\subsection{An example of the network}

It is useful to enumerate the network nodes according to the cluster
structure. First we find \phas- and \fsyn-clusters and enumerate them
with indexes $i\in[0\ldots N_\phc]$ and $j\in [0\ldots N_\fsc]$,
respectively, in ascending order of their sizes, where $N_\phc$ and
$N_\fsc$ are the numbers of corresponding clusters. The index 0
indicates trivial clusters including a single node only. Then the
nodes are assigned the indexes $i_m$ and $j_m$ in accordance to their
membership in clusters, and also the desynchronization degree $\tilde
d_m$ is computed for them, see Eq.~\eqref{eq:phase_desyn}. Now the
real-valued clustering index is defined as
\begin{equation}
  \label{eq:nodord}
  \eta_m=
  \begin{cases}
    -\tilde d_m & \text{if $\tilde d_m>0$}, \\
    i_m+j_m/(N_\fsc+1) & \text{if $\tilde d_m=0$}.
  \end{cases}
\end{equation}
Finally, the nodes are enumerated in the ascending order of $\eta_m$.
The negative $\eta_m$ indicates that the corresponding node is not
synchronized with others, and if in addition $\eta_m$ is very close to
zero the corresponding node is the floating one. The integer part of
positive $\eta_m$ is the index of \phas-cluster to which the node
belongs and the fractional part encodes the \fsyn-cluster index.

\begin{figure*}
  \begin{center}
    \fulfig{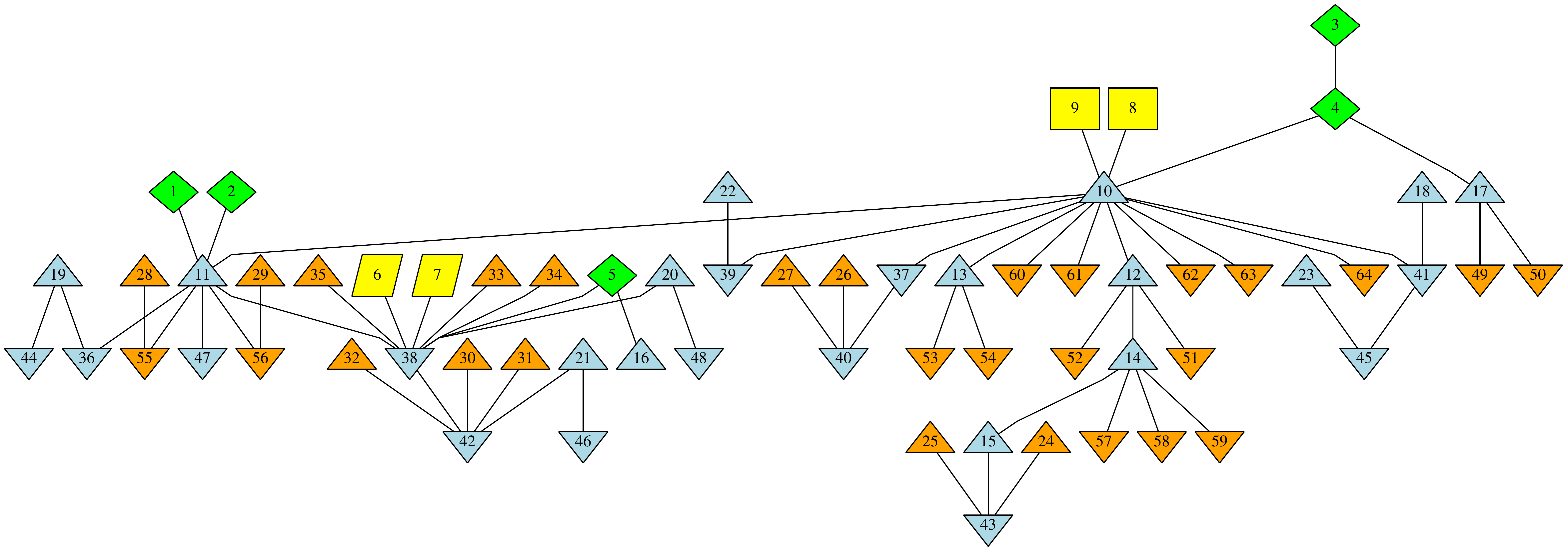}\\
    \begin{tabular}{p{0.3\textwidth}p{0.5\textwidth}p{0.2\textwidth}}
      \raggedright Separated nodes / floating & 1, 2, 3, 4, 5 / 3, 4, 5 &
      green diamonds \tabularnewline
      \hline
      \raggedright Separated \fsyn-clusters & \{6,7\}, \{8,9\} &
      yellow filling \tabularnewline
      \hline
      \raggedright Large \phas-clusters & \{10--35\}, \{36--64\} &
      \raggedright upturned and downturned triangles \tabularnewline
      \hline
      \raggedright \fsyn-clusters embedded into large \phas-clusters & 
      \raggedright \{24,25\}, \{26,27\}, \{28,29\}, \{30,31,32\}, 
      \{33,34,35\}, \{49,50\}, \{51,52\}, \{53,54\}, \{55,56\}, 
      \{57,58,59\}, \{60,61,62,63,64\} & 
      \raggedright orange filling
    \end{tabular}
  \end{center}
  \caption{\label{fig:netw1}An illustration of the clustering of
    network~\eqref{eq:netw_henon} with $N=64$ and $\epsilon=0.17$.
    The nodes and edges represent the connectivity structure and the
    shapes and colors of nodes indicate the states arrived in course
    of the evolution, see the table below the graph. To plot this
    figure we collected data for the cluster identification over
    $\clutim=10^5$ steps.}
\end{figure*}

Figure~\ref{fig:netw1} shows an example of the network structure as
well as its \phas- and \fsyn-clusters emerged in course of the
evolution. The nodes are enumerated according to the ascending order
of $\eta_m$ that is plotted in Fig.~\ref{fig:clv64}(a). The cluster
detection is performed over the whole computation interval $10^5$.

For this particular case there are five nodes that are not
synchronized with others, i.e., have $\eta_m<0$. The first two of them
are essentially separated, $\eta_{1,2}\approx -0.16$, and the nodes 3,
4, and 5 are the floating ones with very small $|\eta_m|$:
$\eta_3=-0.00054$, $\eta_4=-0.00028$, $\eta_5=-2\times 10^{-5}$.

The bulk of nodes form two large \phas-clusters. In our case for these
clusters $3\leq \eta_m<4$ and $4\leq \eta_m<5$, see
Fig.~\ref{fig:clv64}(a). As one can see in Fig.~\ref{fig:netw1}, there
is no any visible relation between the connectivity structure of the
network and the locations of these clusters. The cluster nodes are
mixed so that many nodes of the first cluster are connected with
others only through elements of the second one and vice versa. As we
mentioned above, see Fig.~\ref{fig:sptm}(a,b), the oscillations within
these clusters have opposite phases. Thus, in a wider sense, one can
say that all nodes of these two clusters are phase synchronized, but
some with a phase shift.

Some of nodes of \phas-clusters are synchronized stronger so that they
form \fsyn-clusters embedded into \phas-clusters. For these clusters
$\eta_m$ is fractional and $\eta_m>3$. Observe that all of these
clusters are formed by elements of star-like structures and all
interactions inside \fsyn-clusters pass through hub nodes. The hub
nodes in turn are never synchronized with their subordinate nodes,
see, for example the cluster \{24,25\} connected through a hub
43. Moreover, the hub always belongs to the opposite \phas-cluster:
observe different orientations of the triangles representing the
cluster nodes and the corresponding hubs. This type of synchronization
was first reported in Ref.~\cite{JalanAmritkar2003,*JalanAmritkar2005}
for clusters of phase synchronization. The authors called it driven
synchronization. Later this mechanism was independently described in
Refs.\cite{RemSyn1,RemSyn2} and referred to as remote synchronization.

The structures mentioned so far are typical and always exist for any
$\mat A$ and initial conditions. In some cases, however, like for
example the one shown in Fig.~\ref{fig:netw1}, several more small
\fsyn-clusters appear that are separated from two large
\phas-clusters: the nodes 6 and 7 are fully synchronized with each
other but are not embedded into \phas-clusters. The same is the case
for the nodes 8 and 9.

Finally, notice that remote synchronization can also occur when
``beams'' of a star-like structure include two edges. The nodes 28
and 29 form a \fsyn-cluster, but they can interact only through the
nodes 55 and 56. The latter ones are also synchronized. The opposite
orientation of the corresponding triangles indicates that these
clusters are embedded into different \phas-clusters. This situation
can be treated as remote synchronization of the second order.

\section{\label{sec:tanspc}Structure of the tangent space}

\subsection{The Jacobian matrix}

The Jacobian matrix of the network~\eqref{eq:netw_henon} has a block
form being composed of $N\times N$ matrices:
\begin{equation}
  \label{eq:henon_jac}
  \mat J(t)=
  \begin{pmatrix}
    \mat F(t) & \identmat \\
    \beta\identmat & 0
  \end{pmatrix},
\end{equation}
where 
\begin{equation}
  \label{eq:jac_matr}
  \begin{gathered}
    \mat F(t)=-2\mat G(t)\, [(1-\epsilon)\identmat+\epsilon \mat K^{-1}\mat A],\\
    \mat G(t)=\diagmat\{x_n+\epsilon h_n\},\; \mat K=\diagmat\{k_n\},
  \end{gathered}
\end{equation}
and $\identmat$ is the identity matrix. $\mat J(t)$ has a generic
symplectic structure, i.e., at any $t$ there exists a skew-symmetric
matrix $\mat W(t)$ such that $\mat J(t)\, \mat W(t)\, \mat
J(t)^\transp=-\beta\mat W(t)$. Systems of this type were first
introduced in Ref.~\cite{Dressler88}, however unlike the referenced
paper in our case $\mat W(t)$ is a generic skew-symmetric matrix
depending on $t$:
\begin{equation}
\mat W(t)=
\begin{pmatrix}
  0 & -\mat Q(t) \\
  \mat Q(t) & 0 
\end{pmatrix},
\end{equation}
where $\mat Q(t)$ is a symmetric matrix such that the product $\mat
F(t)\mat Q(t)=\mat M(t)$ is also symmetric. $\mat Q(t)$ can always be
found since any matrix $\mat F(t)$ can always be represented as the
product of two symmetric matrices, $\mat F(t)=\mat M(t)\mat
Q(t)^{-1}$~\cite{BoshFacSymMat}. Due to the this property the Lyapunov
spectrum is symmetric~\cite{Dressler88}:
\begin{equation}
  \label{eq:lyap_sym}
  \lambda_n+\lambda_{N+1-n}=\log \beta 
\end{equation}
The Lyapunov spectra for our system are shown in
Fig.~\ref{fig:lyapful} and discussed below.

\subsection{Pairwise orthogonal eigen-subspaces of the tangent space}

\begin{figure}
  \begin{center}
    \widfig{0.5}{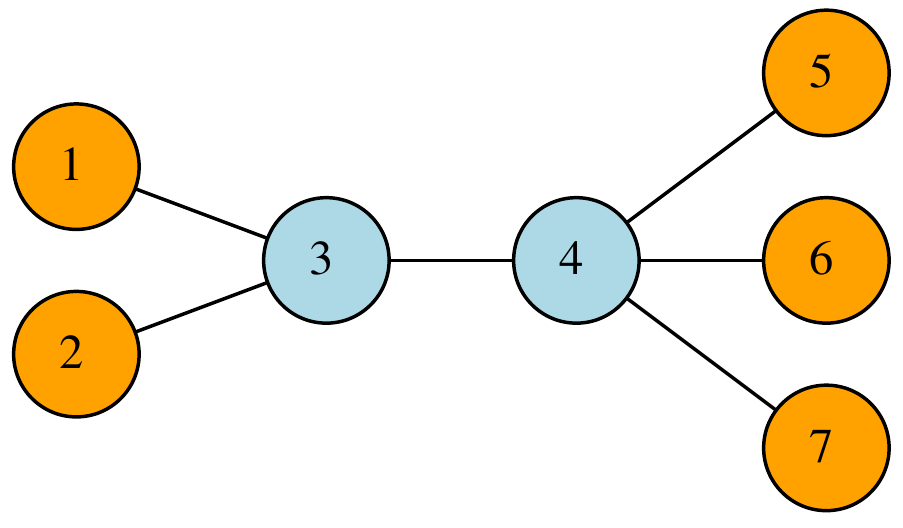}
  \end{center}
  \caption{\label{fig:stars}A toy network with two star-like
    structures. The orange color marks the nodes subjected to the
    remote synchronization.}
\end{figure}

In presence of \fsyn-clusters the tangent space of the
network~\eqref{eq:netw_henon} is split into $N_\fsc+1$ time invariant
subspaces that are pairwise orthogonal, where $N_\fsc$ is the number
of \fsyn-clusters. There are $N_\fsc$ subspaces representing
perturbations transverse to manifolds where the \fsyn-clusters belong,
and the one that includes perturbations longitudinal to all of these
manifolds.

Consider a toy $7\times 7$ network, see Fig.~\ref{fig:stars}. Its
first and second nodes are linked with the third one only forming a
star-like structure and the first \fsyn-cluster. The fifth, sixth and
seventh nodes form the second \fsyn-cluster. The top left block of the
corresponding Jacobian matrix has the form, see
Eq.~\eqref{eq:henon_jac}:
\begin{equation}
  \label{eq:jac_examp}
  \mat F=
  \begin{pmatrix}
    G_1 \epsilon'  & 0 & G_1 \epsilon & 0 & 0 & 0 & 0\\ 
    0 & G_1 \epsilon'  & G_1 \epsilon & 0 & 0 & 0 & 0\\ 
    \frac{1}{3}{g}_{3} \epsilon & \frac{1}{3}{g}_{3} \epsilon &
    {g}_{3} \epsilon'  & \frac{1}{3}{g}_{3} \epsilon & 0 & 0 & 0\\ 
    0 & 0 & \frac{1}{4}{g}_{4} \epsilon & {g}_{4} \epsilon'  &
    \frac{1}{4}{g}_{4} \epsilon & \frac{1}{4}{g}_{4} \epsilon &
    \frac{1}{4}{g}_{4} \epsilon\\ 
    0 & 0 & 0 & G_2 \epsilon & G_2 \epsilon'  & 0 & 0\\ 
    0 & 0 & 0 & G_2 \epsilon & 0 & G_2 \epsilon'  & 0\\ 
    0 & 0 & 0 & G_2 \epsilon & 0 & 0 & G_2 \epsilon' 
  \end{pmatrix}
\end{equation}
where $\epsilon'=1-\epsilon$, $g_i$ are elements of the matrix
($-2\mat G$), see Eq.~\eqref{eq:jac_matr}, and $g_1=g_2=G_1$,
$g_5=g_6=g_7=G_2$ correspond to \fsyn-clusters.

Due to the special form of $\mat F$ there exist vectors of three
types, whose structure is preserved under the mapping with $\mat F$:
\begin{gather}
  \begin{gathered}
    \label{eq:jactl_eigvec0}
    \vec
    v^{\,(0)}=\left(v_1^{(0)},v_2^{(0)},v_3^{(0)},v_4^{(0)},v_5^{(0)},v_6^{(0)},
      v_7^{(0)}\right)^\transp\hspace{-0.5em},\\
    v_1^{(0)}=v_2^{(0)},\; v_5^{(0)}=v_6^{(0)}=v_7^{(0)},
  \end{gathered}\\
  \begin{gathered}
    \label{eq:jactl_eigvec1}
    \vec v^{\,(1)}=\left(v_1^{(1)},v_2^{(1)},0,0,0,0,0\right)^\transp
    \hspace{-0.5em},\\
    v_2^{(1)}+v_1^{(1)}=1,
  \end{gathered}\\
  \begin{gathered}
    \label{eq:jactl_eigvec2}
    \vec
    v^{\,(2)}=\left(0,0,0,0,v_5^{(2)},v_6^{(2)},v_7^{(2)}\right)^\transp
    \hspace{-0.5em},\\
    v_5^{(2)}+v_6^{(2)}+v_7^{(2)}=0.
  \end{gathered}
\end{gather}
The subspaces spanned by these vectors, $\sspace{F}_j=\spanspc\{\vec
v^{\,(j)}\}$, where $j=0,1,2$, are invariant with respect to $\mat F$
and thus form the eigen-subspaces of $\mat F$. Moreover any vector of
the form $\vec v^{\,(1)}$ and $\vec v^{\,(2)}$ is the eigenvector of
$\mat F$ with the eigenvalues $G_{1,2}(1-\epsilon)$. Notice that all
these three subspaces are pairwise orthogonal, i.e., the orthogonal
are any two vectors from these subspaces.

The full Jacobian matrix $\mat J$, see Eq.~\eqref{eq:henon_jac}, also
has three eigen-subspaces $\sspace{J}_j=\spanspc{\{\vec w^{\,(j)}\}}$
spanned by the following block vectors
\begin{equation}
  \vec w^{\,(j)}=
  \begin{pmatrix}
    \vec v_x^{\,(j)}\\
    \vec v_y^{\,(j)}
  \end{pmatrix},
\end{equation}
$j=0,1,2$. Here $\vec v_x^{\,(j)}$ and $\vec v_y^{\,(j)}$ are the
vectors with the
structures~\eqref{eq:jactl_eigvec0}-\eqref{eq:jactl_eigvec2}, related
to perturbations to $x$ and $y$ components of the system. The
dimensions of these subspaces are twice the dimensions of the
eigen-subspaces of $\mat F$. One can find explicitly a couple of
corresponding eigenvectors for subspaces $\sspace{J}_1$ and
$\sspace{J}_2$:
\begin{equation}
  \label{eq:jac_eigvec}
  \vec w^{\,(j)}_{\pm}=\vec v^{\,(j)}
  \begin{pmatrix}
    1 \\
    \beta /\mu^{\,(j)}_\pm
  \end{pmatrix},
\end{equation}
where $\vec v^{\,(j)}$, $j=1,2$, are arbitrary vectors with the
structure \eqref{eq:jactl_eigvec1} and \eqref{eq:jactl_eigvec2},
respectively, and $\mu^{\,(j)}_\pm$ are the corresponding eigenvalues,
\begin{equation}
  \label{eq:jac_eigval}
  \mu^{\,(j)}_\pm=\left(G_j(1-\epsilon)
    \pm\sqrt{G_j^2(1-\epsilon)^2+4\beta}\right)/2.
\end{equation}
For the considered toy network the eigenvalues $\mu^{\,(1)}_+$ and
$\mu^{\,(1)}_-$ both have the multiplicity 1, and the multiplicity of
$\mu^{\,(2)}_+$ and $\mu^{\,(2)}_-$ is 2.

The subspaces $\sspace{J}_1$ and $\sspace{J}_2$ include perturbations
transverse to invariant manifolds of \fsyn-clusters. The dimensions of
these subspaces are 2 and 4, respectively.  All vectors from
$\sspace{J}_0$ contain identical values at sites corresponding to the
same \fsyn-cluster, see Eq.~\eqref{eq:jactl_eigvec0}. It means that
these vectors describe perturbations longitudinal to \fsyn-cluster
manifolds also affecting non cluster nodes. The dimension of
$\sspace{J}_0$ is 8. All three subspaces are orthogonal to each other.

In general case the tangent space of the dynamical network under
consideration is split into a set of eigen-subspaces $\sspace{J}_j$ of
$\mat J$, where $0\leq j\leq N_\fsc$, and $N_\fsc$ is the number of
\fsyn-clusters. These subspaces are time invariant and pairwise
orthogonal.  The subspace $\sspace{J}_j$, where $j\geq 1$, represents
perturbations transverse to the $j$th cluster. It is spanned by
vectors having only $2S_j$ nonzero sites corresponding to $x$ and $y$
variables at cluster nodes, where $S_j$ is the size of the
cluster. Since the sums along $x$ and along $y$ sites have to be zero,
the dimension of this subspace, i.e., the number of independent
vectors, is $2(S_j-1)$.  The subspace $\sspace{J}_0$ is spanned by
vectors of longitudinal perturbations to \fsyn-clusters. These vectors
have identical values at sites corresponding to each node and
independent values at other sites. The dimension of this subspace is
$2(N-M_\fsc+N_\fsc)$, where $M_\fsc$ is the total number of nodes
belonging to all \fsyn-clusters.

\section{\label{sec:nwlfs}Nonwandering localization of 
CLV\lowercase{s} on \fsyn-clusters}

\subsection{The mechanism of localization}

Let $\mat \Gamma(t)$ be a $2N\times 2N$ matrix whose columns are CLVs
at time $t$.  By the definition, this is a unique set of vectors such
that for any $t$ the Jacobian matrix $\mat J(t)$ maps $\mat \Gamma(t)$
to $[\mat C(t+1)\,\Gamma(t+1)]$, where $\mat C(t)$ is a diagonal
matrix logarithms of whose elements are finite time Lyapunov
exponents~\cite{CLV2012}. In the other words, the tangent space
operator, that is $\mat J$ for discrete time systems, maps each CLV at
$t$ to the stretched or contracted CLV at $t+1$.

The direct sum of the subspaces $\sspace{J}_j$, $0\leq j\leq N_\fsc$,
is equal to the whole tangent space, the subspaces are time invariant
and moreover pairwise orthogonal. Thus each of them holds a set of
CLVs related to perturbations to individual clusters or to non-cluster
nodes. The number of these vectors is equal to the dimension of the
corresponding subspace $\sspace{J}_j$. These CLVs can freely evolve
only within their subspaces and never leave them. Let us assume that
this is not the case and there exists a probe CLV not fully belonging
to one of the subspaces $\sspace{J}_j$. This vector can always be
decomposed into a linear combination of vectors from
$\sspace{J}_j$. In course of the evolution the vectors of this
decomposition grow or decay exponentially, on average, but always stay
within their subspaces. The rates of this growth or decay are the
Lyapunov exponents. One of the vectors with the largest Lyapunov
exponent will always dominate all others so that our probe CLV will
fall into the corresponding subspace. Thus each CLV indeed belongs to
one of $\sspace{J}_j$. In principle, however, the Lyapunov exponents
from different subspaces can coincide. In this case the corresponding
CLVs will be linear combinations of vectors from these subspaces.

The CLVs related to transverse perturbations of \fsyn-clusters have
nonzero elements only at sites corresponding to the cluster
nodes. Since the considered \fsyn-clusters are small, the
corresponding CLVs are highly localized. Moreover, this localization
is nonwandering, i.e., the nonzero vector elements always have a fixed
location. 

Localization of CLVs is a well known phenomenon. However for
chain-like systems whose nodes have identical pattern of connections
the localization sites wander around irregularly from node to
node~\cite{PikKur94,*PikPol98,CLVPazo}. The nonwandering localization
of CLVs is known to occur due to the inhomogeneous structure of a
system. It was already reported for a disordered medium in
Ref.~\cite{SzendLopRodr08}. From a general point of view the
nonwandering localization of CLVs in our system also occurs because
the system is highly inhomogeneous, namely, due to the star-like
structures when there are highly connected hubs and low connected
subordinate nodes.

\subsection{Defects of Lyapunov spectra}

\begin{figure}
  \begin{center}
    \widfig{1}{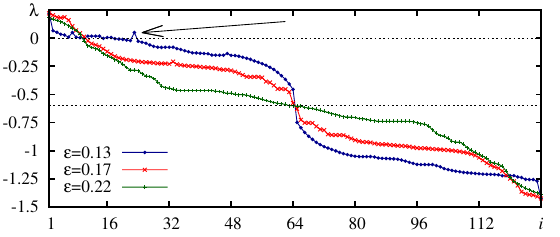}
  \end{center}
  \caption{\label{fig:lyapful}(color online) Lyapunov spectra of the
    network~\eqref{eq:netw_henon} with various coupling strengths,
    $\epsilon=0.13$, $0.17$, and $0.22$. The upper dotted line marks
    zero, while the lower one is the symmetry axis at $(\log
    \beta)/2$. The arrow points an example of anomalous
    behaviour. $N=64$.}
\end{figure}

Let us consider the Lyapunov spectra of the
network~\eqref{eq:netw_henon}, see Fig.~\ref{fig:lyapful}. Observe the
symmetry of the curves, emerging due to the generic symplectic
structure of the Jacobian matrix, see Eq.~\eqref{eq:lyap_sym}. The
theory behind the algorithm for Lyapunov
exponents~\cite{Benettin,Shimada79} is based on the hierarchy of
domination of tangent vectors obeyed by different Lyapunov
exponents. During the computation we evolve a set of tangent vectors
mapping them with the Jacobian matrix and thus allowing to align along
the most expanding available directions. To exclude the alignment of
all the vectors along the same directions, we periodically
orthogonalize them. So the first one points the most expanding
direction, the second one, as well as all others, are orthogonal to it
and can only align along the second expanding direction and so on. The
average exponential growth rates of these vectors are the Lyapunov
exponents. Obviously they have to appear a non-ascending order.

However, in our case the non-ascending order can be broken, see the
arrow in Fig.~\ref{fig:lyapful}. Notice the absence of the symmetrical
defect on the second part of the spectrum. This abnormal behaviour is
related to the splitting of the tangent space into the orthogonal
subspaces $\sspace{J}_j$. Right after the start of the iterations, the
tangent vectors have random directions. If the local expansion rates
for some of the subspaces $\sspace{J}_j$ highly deviate from the
corresponding Lyapunov exponents, this subspace can attract wrong
vectors. In ``normal'' situation the wrong orientation of vectors is
fixed after a transient time when the influence of local rates
decays. But in our case, since the subspaces $\sspace{J}_j$ are time
invariant, the vectors can be trapped within inappropriate
subspaces. As a result we observe the broken order of Lyapunov
exponents as pointed by the arrow in Fig.~\ref{fig:lyapful}.

One can try to avoid this trapping by adding a small noise to tangent
vectors after each iteration. The noise is expected to push out the
vectors from their traps giving them a chance to arrive at the
appropriate subspace. Our tests showed that even very small noise of
the order $10^{-10}$ can smoothen the defects of Lyapunov
spectra. However, instead of the pushing out of the trapped vectors,
the noise destroys the splitting of the tangent space at all. The
vectors do not gain the structures described by
Eqs.~\eqref{eq:jactl_eigvec0}-~\eqref{eq:jactl_eigvec2} any more. Thus
this is inappropriate approach since the existence of the tangent
subspaces $\sspace{J}_j$ is one of the essential features of our
system.

\begin{figure}
  \begin{center}
    \widfig{1}{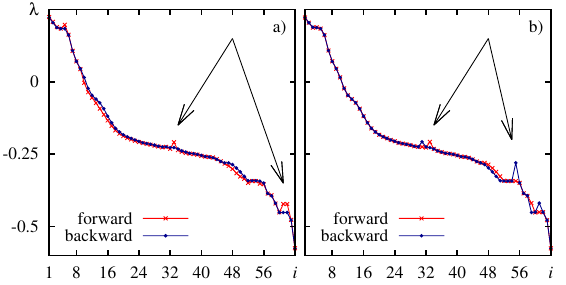}
  \end{center}
  \caption{\label{fig:lyflyb} Lyapunov spectra computed in parallel
    with computation of CLVs via (a) IR- and (b) LU-methods. Two
    curves in the panels correspond to the exponents computed in
    course of forward- and backward-time stages. The arrows point the
    areas of essential deviations of the curves from each other.}
\end{figure}

\subsection{Structure of CLVs}

Now we turn to the CLVs. There are two numerical methods for computing
CLVs whose ideas where published simultaneously. The method
reported in Ref.~\cite{CLVGin} shall be referred below as
IR-method. It computes CLVs in course of iterations backward in time
with inverted upper triangle matrices $\mat R$ previously obtained on
the forward-time stage as a result of so called QR matrix
decompositions. The other method first reported in Ref.~\cite{CLVWS}
was later improved in Ref.~\cite{CLVPazo} and then it was reformulated
in a more efficient form in Ref.~\cite{CLV2012}. This method shall be
referred as LU-method since it computes CLVs as a result of LU
decomposition of matrices of scalar products of orthogonal Lyapunov
vectors computed in course of forward- and backward-time procedures.

Both of the methods for CLVs includes the iterations with tangent
vectors forward and backward in time. To compute CLVs correctly these
iterations have to provide the identical orderings of tangent vectors,
even if this does not correspond to the non-ascending order of the
Lyapunov exponents. Unfortunately the trapping of vectors within
inappropriate subspaces $\sspace{J}_j$ can occur independently and
thus differently on forward and backward stages. These situations can
be identified by comparing Lyapunov exponents computed in parallel
with forward and backward stages, see Fig.~\ref{fig:lyflyb}. One can
see that besides natural small and smooth deviations, related to an
unavoidable numerical noise, there are points marked by arrows where
the orders of the exponents do not coincide. It indicates that the
forward- and backward-time data do not exactly match so that the
corresponding CLVs are not quite correct. These abnormal deviations of
the curves are found to be is less pronounced for the IR method, and
below we shall use it to for computing CLVs.

\begin{figure}
  \centering \widfig{1}{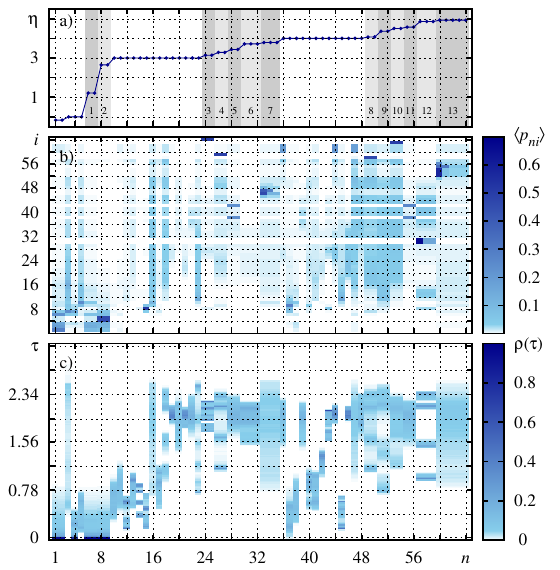}
  \caption{\label{fig:clv64}(color online) (a) Clustering index
    $\eta_n$, see Eq.~\eqref{eq:nodord}. The nodes are enumerated
    according to the ascending order of $\eta_n$. Grey labeled stripes
    indicate \fsyn-clusters.  (b) Average node related CLVs. (c)
    Distributions of $\tau_n$, see Eq.~\eqref{eq:tau}. For all panel
    $n$ is node number and $i$ is the vector number. $N=64$,
    $\epsilon=0.17$. The matrix $\mat A$ and initial conditions are
    the same as in Fig.~\ref{fig:netw1}.}
\end{figure}

Figure~\ref{fig:clv64}(b) shows CLVs averaged in time. Since two
variables are associated with each node, we consider the node related
CLVs $\clvpn_{ni}=\gamma_{2n-1,i}^2+\gamma_{2n,i}^2$, where
$\gamma_{ji}$ is the $j$the element of the $i$th CLV,
$i,j=1,\ldots,2N$, $n=1,\ldots,N$.  Because each CLV has a unit
length, $\sum_{n=1}^N \clvpn_{ni}=1$ for any
$i$. Figure~\ref{fig:clv64} corresponds to the network shown in
Fig.~\ref{fig:netw1}. The nodes of the network are enumerated
according to the ascending order of $\eta_n$, see
Eq.~\eqref{eq:nodord}. The curve $\eta_n$ is shown in
Fig.~\ref{fig:clv64}(a). Grey stripes in this panel mark
\fsyn-clusters.

According the discussion above, there are CLVs localized of
\fsyn-clusters. The most clear examples correspond to the clusters
3,4,7,8,10,12, and 13. The number of vectors has to be one less then
the number of nodes in the cluster (notice that only the first part of
the symmetric spectrum is shown, one more set of vectors also exist in
the second part). Thus each of two-node clusters 3,4,8,and 10 produces
a single localized CLV. The three-node clusters 7 and 12 generate
pairs of CLVs. Finally, the five-node cluster 13 are characterized by
four CLVs.

The two-node clusters 5 and 11 generate two CLVs, localized
simultaneously on both of these cluster. These clusters includes the
nodes \{28, 29\} and \{55,56\}, respectively. As we already discussed
above, they demonstrate remote synchronization of the second order,
since the nodes 28 and 29 are synchronized through the nodes 55 and
56, see Fig.~\ref{fig:netw1}. Due to this reason the exponential
growth rates in the subspaces corresponding to these two clusters are
always identical and no one of them dominates. The resulting CLVs are
linear combinations of vectors localized on these clusters.

The clusters 1, 2, 6, and 9 are problematic. The two-node cluster 2
has two localized CLV instead of the expected one, and the clusters
1,6, and 9 do not have any clearly localized CLVs. We address this
issues to the fails of the numerical methods due to the trapping of
tangent vectors within inappropriate subspaces $\sspace{J}_j$, see the
discussion above.

All CLVs not localized on \fsyn-clusters belong to $\sspace{J}_0$
representing longitudinal perturbations to these clusters. It means
that they have to have identical values at sites corresponding to
\fsyn-clusters. One can see that this requirement is fulfilled well
even for problematic clusters.

\section{\label{sec:nwlph}Nonwandering localization of
  CLV\lowercase{s} on nodes separated from \phas-clusters}

\subsection{Properties of localized vectors}

Besides the localization on \fsyn-clusters one can also observe in
Fig.~\ref{fig:clv64}(b) that the first six vectors are localized on
nodes 1, 2, 6-9. The common property of these nodes is that they do
not belong to \phas-clusters, see Fig.~\ref{fig:netw1}. 

To clarify it we shall detect the clusters at $\clutim=20$. Since the
oscillations of phase synchronized nodes are very close to periodic
with the period 2, see Fig.~\ref{fig:sptm}, this short $\clutim$ is
the smallest reasonable value required to identify intermittent
attachments and detachments of nodes to \phas-clusters. Running over
the computation interval and performing serial detections of
\phas-clusters we assign to each node at each time step a flag
signalling whether this node belongs to a \phas-cluster or not. Also
we compute CLVs and for each vector at each time step using the flags
we find a sum
\begin{equation}
  \label{eq:seploc}
  p_\spr(t)=\sum_{n=1}^{M_\spr(t)} \clvpn_{ni}(t),
\end{equation}
where $M_\spr(t)$ is the number of nodes separated from the
\phas-clusters. The nodes in this equations are assumed to be
enumerate in a such a way that the separated nodes go first. Since
$\sum_{n=1}^N \clvpn_{ni}=1$, $p_\spr$ indicates what a fraction of
nonzero CLV elements belongs to the separated nodes. The upper limit
$p_\spr=1$ tells that all nonzero CLV elements are localized on
separated nodes, while $p_\spr=0$ shows that all nonzero CLV elements
are localized on \phas-clusters.

\begin{figure}
  \centering \widfig{1}{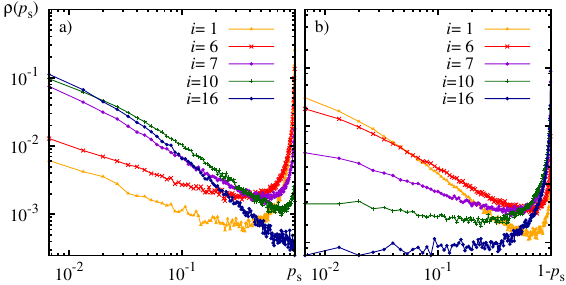}
  \caption{\label{fig:seploc_ps1}(color online) Power law decays of
    $\rho(p_\spr)$ (a) near $p_\spr=0$, and (b) near
    $p_\spr=1$. Double logarithmic scales are used for both axis. The
    vector numbers are shown in the legends. The matrix $\mat A$ and
    initial conditions are the same as in Figs.~\ref{fig:netw1} and
    \ref{fig:clv64}.}
\end{figure}

\begin{figure}
  \centering \widfig{1}{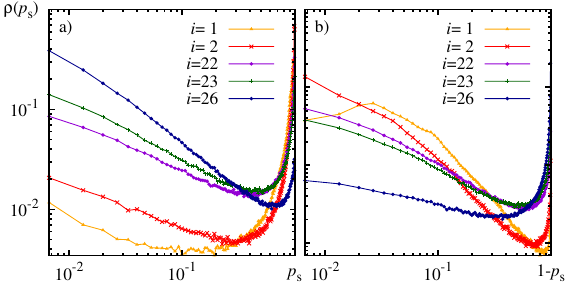}
  \caption{\label{fig:seploc_ps2}(color online) Distributions
    $\rho(p_\spr)$ at $N=128$, $\epsilon=0.22$.}
\end{figure}

The distributions of $p_\spr$ are found to have two maxima, one at
$p_\spr=0$ and the other at $p_\spr=1$, and they decay fast towards to
the middle area.  Figure~\ref{fig:seploc_ps1} plotted for the network
in Figs.~\ref{fig:netw1} and \ref{fig:clv64} shows that the decays
near both edges are obeyed to power laws. Notice that the orders of
the curves representing different vectors are different at the left
and right edges. For the vector $i=1$ $\rho(0)<\rho(1)$. It means that
this vector is preferably localized on nodes not attached to the
\phas-clusters. This is also the case for all vectors up to the sixth
one, while for the seventh vector we observe
$\rho(0)>\rho(1)$. Staring from this vector all other CLVs are
preferably localized on the nodes attached to the \phas-clusters.

Figure~\ref{fig:seploc_ps2} shows the distributions $\rho(p_\spr)$ at
$N=128$ and $\epsilon=0.22$. One again observes the power laws near
the edges and preferable localization of the vectors $1\leq i\leq 22$
on the nodes not attached to the \phas-clusters, since for these
vectors $\rho(0)<\rho(1)$. Also notice the essential deviation from
the power law of the distribution for $i=1$ near the right edge, see
Fig.~\ref{fig:seploc_ps2}(b). This is the result of the approaching of
$\epsilon$ to the right boundary of the area of our consideration
marked in Figs.~\ref{fig:lyeps} and \ref{fig:clustering}. We tested
more distributions at $\epsilon=0.24$ and observed that the deviation
from the power law near $p_\spr=1$ gets higher. But nevertheless we
still can distinguish the CLVs localized on separated nodes by
comparing the edge values of the distributions $\rho(0)$ and
$\rho(1)$.

\begin{figure}
  \centering \widfig{1}{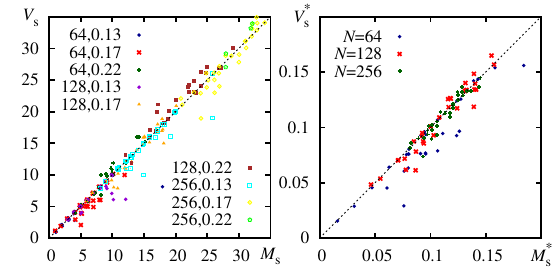}
  \caption{\label{fig:seplocnum}(color online) (a) The number of nodes
    non-synchronized with the \phas-clusters $M_\spr$ vs. the number
    of CLVs $V_\spr$ localized on them. The legend shows the values of
    $N=64$, $128$ and $256$ and $\epsilon=0.13$, $0.13$, and
    $0.22$. (b) Rescaled values $M^*_\spr$ vs. $V^*_\spr$, see
    Eq.~\eqref{eq:scale_v} for $N=64$, $128$ and $256$ at
    $\epsilon=0.17$.}
\end{figure}

Thus the first CLVs, whose number we denote as $V_\spr$, are
preferably localized on the nodes not synchronized with the
\phas-clusters. Notice that due to the symmetry there are more
$V_\spr$ localized vectors in the opposite end of the spectrum.

Since these CLVs have nonzero values mainly on a limited and permanent
set of nodes whose number we denote as $M_\spr$, the number of these
vectors have to be at least approximately equal to the number of these
nodes, $V_\spr\approx M_\spr$. To verify it we generate different
matrices $\mat A$ and find the separated nodes for it. Then we find
CLVs and compute the relative frequency $P(p_\spr>0.5)$, where
$p_\spr$ is computed as discussed above, see
Eq.~\eqref{eq:seploc}. The vector is treated as localized on the
separated nodes when $P>0.5$. The number of such vectors $V_\spr$ as a
function of the number of separated nodes $M_\spr$ is plotted in
Fig.~\ref{fig:seplocnum}. Since $M_\spr$ and $V_\spr$ are integer, the
points of the plot will overlap each other. To avoid it and show the
areas where the points fall more often as dense clouds we add random
numbers $\xi\in (-0.2,0.2)$ to data: $M_\spr+\xi$ and $V_\spr+\xi$.
Panel (a) shows nine data sets computed at $\epsilon=0.13$, $0.17$,
and $0.22$ for $N=64$, $128$, and $256$. The points are fitted very
well by the straight line $V_\spr=M_\spr$ that confirms the expected
relation between the number of localized vectors and the number of
separated nodes.

Figure~\ref{fig:seplocnum}(b) illustrates the scaling of $M_\spr$ and
$V_\spr$ with the network size $N$. One sees that though different
matrices $\mat A$ result in different $M_\spr$ and $V_\spr$, the
scaling
\begin{equation}
  \label{eq:scale_v}
  M^*_\spr=M_\spr/N,\; V^*_\spr=V_\spr/N,
\end{equation}
results in the gathering of points withing the same ranges. It means
that the number of nodes separated from the \phas-clusters as well
as the number of localized on them CLVs grow with $N$ as $M_\spr\sim
N$, and $V_\spr\sim N$. Notice that this agrees with previously
discussed scaling of the number of nodes attached to the
\phas-clusters, see Eq.~\eqref{eq:scale_m}.

We also checked the signs of Lyapunov exponents corresponding to the
localized CLVs. In all cases the localized CLVs had positive Lyapunov
exponents and the total number of positive Lyapunov exponents was
always higher then $V_\spr$.

\subsection{Properties of localization nodes}

The separated nodes where the first CLVs are localized have common
specific feature related to the instantaneous square deviations of a
node from its neighborhood:
\begin{equation}
  \label{eq:tau}
  \tau_n(t)=h_n^2(t).
\end{equation}
where $h_n(t)$ is given by
Eq.~\eqref{eq:gn}. Figure~\ref{fig:clv64}(c) shows the distributions
of $\tau_n$. One can see that the distributions at nodes 1, 2, 6-9
have the maximum in zero and they decay monotonically. On contrary,
the distributions at nodes with numbers $n\geq 10$ are quite
different: all of them are separated from zero, and in some cases they
are multimodal.

Since nodes 3, 4, and 5 are the floating ones, as indicate
corresponding values of $\eta_n$ in Fig.~\ref{fig:clv64}(a), the forms
of corresponding distributions of $\tau_n$ are ambiguous. On the one
hand side, the distribution at node 4 looks as in non-floating
ones. However, the distributions in nodes 3 and 5 correspond to the
situation when a node belong to a \phas-cluster at $n\geq 10$.

This can be clarified by finding the clusters at short interval,
$\clutim=20$. Performing the serial cluster detections with this
$\clutim$ for the network in Figs.~\ref{fig:netw1} and \ref{fig:clv64}
we found that the separated nodes 1 and 2 as well as the nodes of the
small \fsyn-clusters 6-9 can sometimes be attached to a \phas-cluster,
but approximately 90\% of time they oscillate separately. Contrary to
this the floating node 3 is not synchronized with the \phas-clusters
only 0.0008\% of time steps, and node 5 is separated 0.0002\% of
time. However the node 4 remains separated from \phas-clusters during
0.0022\% of time steps. Though this is still a very small value but it
is one order higher then for the nodes 3 and 5. Thus the form of the
distribution of $\tau_n$ depends on the percentage of time that the
node spends being not synchronized with the \phas-clusters.

\begin{figure}
  \centering \widfig{1}{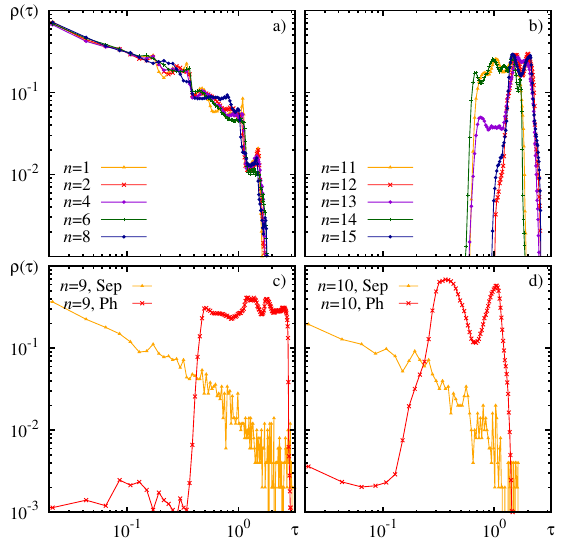}
  \caption{\label{fig:seploc_tau}(color online) Distributions
    $\rho(\tau)$ at $N=128$, $\epsilon=0.13$ at different nodes
    $n$. The nodes are enumerated according to the growth of $\eta_n$,
    see Eq.~\eqref{eq:nodord}. Panel (a) shows the separated nodes,
    while panel (b) corresponds to the nodes attached to the
    \phas-clusters. Panels (c) and (d) demonstrate distributions at
    the floating nodes $n=9$ and $10$, respectively, computed
    independently when the node is separated (label ``Sep'' in the
    legend) and attached to the \phas-clusters (label ``Ph'').}
\end{figure}

Figure~\ref{fig:seploc_tau} exemplifies the typical forms of the
distributions of $\tau_n$ in more detail. A network generated to plot
this figure had eight purely separated nodes and two floating
ones. The nodes are assumed to be enumerated according to the growth
of clustering index $\eta_m$, see Eq.~\eqref{eq:nodord}. In panel (a)
one can see that the distributions of $\tau_n$ at the separated nodes
have power law shape near the origin, right after that it decays to
zero, and moreover the shapes of distributions in all of these nodes
are almost identical. On contrary, the distributions at \phas-cluster
nodes are well separated from zero and can have multiple maxima, see
panel (b). To plot the distribution for floating nodes in panels (c)
and (d) we collected the data in two arrays, one was used when the
node was attached to a \phas-cluster, and the other when it was
separated. One can see that oscillating separately the floating node
demonstrate the power law distribution of $\tau_n$. The exponent
coincides with the exponents of the distributions for purely separated
nodes, cf. the slopes of the curves in panels (a) with the slopes of
the corresponding curves in panels (c) and (d). When the floating node
is attached to a \phas-cluster its distribution corresponds in bulk to
the distributions at purely cluster nodes, cf. the curves in panel (b)
with the corresponding curves in panels (c) and (d). However a remnant
power law tail near the origin can also be observed in panel (d).

One can see in Fig.~\ref{fig:netw1} that each of the separated nodes
where the first CLVs are localized has only one connection. This is
typical for the localization nodes. Computing the connectivity degrees
$k_n$ in parallel with the data for Fig.~\ref{fig:seplocnum} we found
that in the most cases $k_n=1$ though rarely it can be
higher. Nevertheless, the average connectivity degree of the separated
nodes where CLVs are localized is less then 2.

Altogether, the first $V_\spr$ CLVs are localized on $M_\spr$
nodes. These nodes have specific properties: they are not synchronized
with large \phas-clusters, in most cases they have only one
connection, and the distributions of $\tau_n$ at these nodes have
identical power law shapes. The core set of these nodes remains
unchanged in course of the dynamics (however there can exist a few so
called floating nodes). It means that this localization of CLVs is
nonwandering. Since the localization nodes can be found without the
straightforward computation of CLVs, we can predict where the first
$V_\spr$ CLVs are localized.

\section{\label{sec:summ}Summary and conclusion} 

In this paper we found that CLVs for a dynamical network can
demonstrate nonwandering localization on nodes that can be found
without the computation of CLVs. This is an example of explicit
relations between dynamics of a system and the associated tangent
space dynamics.

Random scale-free dynamical networks of H\'{e}non maps are
considered. The networks are generated using preferential attachment
mechanism, and the resulting network always have $N$ nodes and $N-1$
connections.

The dynamics of such network is chaotic. Though the synchronization of
the whole network is not observed, the nodes can form synchronized
clusters. Full chaotic synchronization as well as phase
synchronization are possible. The number of clusters depend on the
coupling strength. We limit ourselves with a range of coupling
strengths were there are two large phase clusters including together
almost all nodes, and many small fully synchronized clusters. Most of
the them are embedded into the phase clusters while several ones can
be separated.

Due to the presence of clusters, covariant Lyapunov vectors are found
to be localized. Each cluster of $S_\fsc$ fully synchronized nodes is
associated with $2(S_\fsc-1)$ covariant vectors all of whose sites are
strictly zeros except for the nodes corresponding to the
clusters. This localization is nonwandering and predictable since we
can find nonzero vector sites without computing the covariant
vectors. However it is unclear which vector will be localized on the
particular cluster.

One more mechanism of localization is related to the phase
clusters. The first $V_\spr$ CLVs are localized on $M_\spr$ nodes that
oscillate separately from the phase clusters. This localization is not
quite strict as the previous one, and the vectors can have nonzero
sites on nodes attached to the phase clusters. But the probability of
localization on separated nodes is always higher and this is the
criterion for distinguishing of these vectors. The number of vectors
$V_\spr$ and the number of separated nodes $M_\spr$ are equal, however
since the localization is not strict this equality is approximate. As
well as the localization of clusters of full synchronization this is
the nonwandering and predictable localization. Finding the nodes
oscillating separately from the phase clusters we can say in advance
where the first CLVs will be preferably localized and what will be
their number. The nodes of localization have specific features: they
are very low connected (only one connection, in the most cases), and
they demonstrate identical power law distributions of square
deviations of dynamical variables from their neighborhood.

The a priori knowledge about the localization of the covariant vectors
opens perspectives of wider utilizing of these vectors. By the
definition these vectors show how the development of perturbations
occurs. When the locations of areas of the most intensive development
is permanent and predictable, the interesting problem arises to
organize an effective low energy forcing to the system using this
areas.

Computing CLVs for the dynamical networks with full synchronization
clusters we found that both known methods can be not quite correct due
the splitting of the tangent space into a set of time invariant
pairwise orthogonal subspaces. In view of the great interest of
researcher to the dynamical networks a challenging task emerges to
modify the numerical methods for CLVs to fix this problem.

PK thank U. Parlitz for stimulating discussions.

PK acknowledges the President RF program of support of leading Russian
research schools NSh-1726.2014.2.

\bibliography{nwl}

\end{document}